\documentclass[proceedings]{JHEP3} 
\PrHEP{}
\conference{International Workshop on Astroparticle and High Energy
Physics}
\usepackage{epsfig}                   
\def\baa             {\begin{array}}
\def\eaa             {\end{array}}
\def\beqaa           {\begin{eqnarray}}
\def\eeqaa           {\end{eqnarray}}
\def\beqaad          {\begin{eqnarray*}}
\def\eeqaad          {\end{eqnarray*}}
\def\bce             {\begin{center}}
\def\ece             {\end{center}}
\def\btabu           {\begin{tabular}}
\def\etabu           {\end{tabular}}

\def\ti              {\tilde}

\def\st              {\ti t}
\def\sb              {\ti b}

\def\sl              {\ti \ell}

\newcommand{\gsim}{\;\raisebox{-0.9ex}
           {$\textstyle\stackrel{\textstyle >}{\sim}$}\;}
\newcommand{\lsim}{\;\raisebox{-0.9ex}
           {$\textstyle\stackrel{\textstyle<}{\sim}$}\;}

\def\ti{\tilde}

\def\lsim{\raise0.3ex\hbox{$\;<$\kern-0.75em\raise-1.1ex\hbox{$\sim\;$}}}
\def\gsim{\raise0.3ex\hbox{$\;>$\kern-0.75em\raise-1.1ex\hbox{$\sim\;$}}}

\title{Impact of CP phases on SUSY particle production and decays}

\author{\speaker{Alfred Bartl} \\
        Institut f\"ur Theoretische Physik, Universit\"at Wien, A-1090 Vienna\\
        E-mail: \email{bartl@ap.univie.ac.at}}

\author{Hans Fraas\\
        Institut f\"ur Theoretische Physik und Astrophysik, 
        Universit\"at W\"urzburg,\\
        D-97074 W\"urzburg}

\author{Stefan Hesselbach\\
        Institut f\"ur Theoretische Physik, Universit\"at Wien, A-1090 Vienna}

\author{Keisyo Hidaka\\
        Department of Physics, Tokyo Gakugei University, Koganei, 
        Tokyo  184-8501, Japan}

\author{Thomas Kernreiter\\
      Institut f\"ur Theoretische Physik, Universit\"at Wien, A-1090 Vienna}
       
\author{Olaf Kittel\\
       Institut f\"ur Theoretische Physik und Astrophysik,
       Universit\"at W\"urzburg,\\ 
       D-97074 W\"urzburg and\\
       IFIC - Instituto de F\'{\i}sica Corpuscular,
       Universitat de Val{\`e}ncia,\\
       Edificio Institutos d'Investigaci{\'o},
       E-46071 Val{\`e}ncia, Spain}

\author{Werner Porod\\
        Institut f\"ur Theoretische Physik, Universit\"at Z\"urich,
  CH-8057 Z\"urich}

\abstract{We report on the results of a phenomenological study of 
top squarks $(\ti t_{1,2})$ and
bottom squarks $(\ti b_{1,2})$ in the Minimal Supersymmetric
Standard Model (MSSM) with complex parameters $A_t, A_b, \mu$ and $M_1$.
In particular we focus on the CP phase dependence
of the branching ratios of $\tilde t_{1,2}$ and $\tilde b_{1,2}$ decays.
We find that the effect of the phases on the $\tilde t_{1,2}$ and
$\tilde b_{1,2}$ decays can be quite significant in a large region
of the MSSM parameter space. We also study a CP asymmetry in sfermion decays.}

\dedicated{\rm IFIC/03-59, TGU-33, UWThPh-2003-43, WUE-ITP-2003-026, 
  ZU-TH 21/03, hep-ph/0312306}

\begin{document}
\section{Introduction}

Supersymmetry (SUSY) is one of the best studied extensions 
\cite{susy} of the Standard Model (SM). 
SUSY gives us also the benefit of introducing potential new
sources of CP violation \cite{Dugan:1984qf}.
As the small amount of CP violation
in the SM is not sufficient to explain the baryon asymmetry
of the universe, it is necessary to study 
all implications of the
complex SUSY parameters.
The search for SUSY 
will be one of the main goals of all future colliders.
An $e^+e^-$ linear collider will be an ideal machine for the determination of
the underlying SUSY parameters \cite{tdr}.

In this talk we present the results of our studies 
\cite{stosbo,short,Bartl:2003ck} of the effects
of complex SUSY parameters on the phenomenology of
the scalar top quark and scalar bottom quark system.  
Analysing the properties of 3rd generation sfermions
is particularly interesting, because of the effects of
the large Yukawa couplings.
The lighter sfermion mass eigenstates may be among the
light SUSY particles and they could be investigated 
at
$e^+ e^-$ linear colliders
\cite{sfLCnojiri,sfLCbartl}.
Previous analyses of the decays of the
3rd generation sfermions $\tilde{t}_{1,2}$, $\tilde{b}_{1,2}$,
$\tilde{\tau}_{1,2}$ and $\tilde{\nu}_\tau$ in the 
Minimal Supersymmetric Standard Model (MSSM) with real
parameters have been performed in 
Refs.~\cite{stop2-stau2}--\cite{eberlRealSf}. 

In the MSSM the SUSY 
parameters $A_f$, $\mu$ and $M_i$  ($i=1,2,3$) are in general 
complex, where $A_f$ is the trilinear scalar coupling parameter
of the sfermion $\tilde{f}_i$, 
$\mu$ is the Higgs-higgsino mass parameter 
and $M_1$, $M_2$ and $M_3$ are 
the U(1), SU(2) and SU(3) gaugino mass parameters, respectively.
We will first study the phase dependence of the  
decay branching ratios of the top squarks and bottom squarks,
which are CP-even observables. As we will show, the phase 
dependence of these decay branching is indeed
suitable to obtain informations about the SUSY CP phases.
The situation is quite similar to that of the third generation slepton 
system, where the decay branching ratios of the staus 
$\tilde{\tau}_{1,2}$ and $\tau$-sneutrino
$\tilde{\nu}_\tau$ can be used to get information
on the phases of the stau
and gaugino-higgsino sectors \cite{CPslepton}.
In our study we
will use the MSSM as a general framework and assume that the
parameters $A_t$, $A_b$, $\mu$ and $M_1$ have 
the phases $\varphi_{A_t}$, $\varphi_{A_b}$, $\varphi_\mu$ and
$\varphi_\mathrm{U(1)}$, respectively (taking $M_{2,3}$ real).
We take into account explicit CP violation in the Higgs sector
\cite{Demir:1999hj,ref2,feynhiggs}.
Furthermore, we also take into account the constraints on the SUSY 
parameters which follow from the experimental data on the 
rare decay $b \rightarrow s \gamma$ \cite{bsgamma}.

We will also consider a CP-odd observable in sfermion decays, 
which provides a more direct signal for the presence of CP phases.
This observable is a CP-sensitive asymmetry which follows from 
triple product correlations \cite{Bartl:2003ck,Bartl:2002hi}.

\section{Decay Branching Ratios of Top Squarks and Bottom Squarks}

Considering first top squark and bottom squark production, 
the reaction 
$e^+ e^- \to \tilde{q}_i \bar{\tilde{q}}_j$, 
$\tilde{q}_i = \tilde{t}_i, \tilde{b}_i$, 
proceeds via
$\gamma$ and $Z$ exchange in the $s$-channel. 
The tree-level cross sections \cite{sfLCbartl} of these
reactions do not explicitly depend on the phases $\varphi_\mu$ 
and $\varphi_{A_q}$, because the 
$Z \tilde{q}_i \tilde{q}_i$ couplings are real and in 
$e^+ e^- \to \tilde{q}_1 \bar{\tilde{q}}_2$ only $Z$ exchange
contributes. The cross sections depend only on the mass eigenvalues
$m_{\tilde{q}_{1,2}}$ and on the mixing angles
$\cos^2\theta_{\tilde{q}}$. Therefore, they depend only implicitly on the
phases via the $\cos(\varphi_\mu + \varphi_{A_q})$ dependence of
$m_{\tilde{q}_{1,2}}$ and $\theta_{\tilde{q}}$.

In the following we will present numerical results for the phase
dependences of the $\st_i$ and $\sb_i$ partial decay widths and branching
ratios. We will treat the fermionic decays
\begin{equation}
\tilde{q}_i \to q' + \tilde{\chi}^\pm_k, i=1,2, k=1,2
 \label{eq:gamtC}
\end{equation}
and
\begin{equation}
\tilde{q}_i \to q + \tilde{\chi}^0_k, i=1,2, k=1,..,4
 \label{eq:gamtN}
\end{equation}
and the bosonic decays
\begin{equation}
\tilde{q}_i \to W^\pm + \tilde{q}'_j, i=1,2, j=1,2, j \leq i,
\end{equation}
\begin{equation}
\tilde{q}_i \to H^\pm + \tilde{q}'_j, i=1,2, j=1,2, j \leq i,
\end{equation}
\begin{equation}
\tilde{q}_2 \to Z + \tilde{q}_1,
\end{equation}

and
\begin{equation}
\tilde{q}_2 \to H_i + \tilde{q}_1, i= 1,2,3.
\end{equation}

These partial decay widths depend on the SUSY parameters of the squark system 
$M_{\tilde{Q}}$, $M_{\tilde{U}}$, $M_{\tilde{D}}$, $\tan\beta$, 
$|\mu|$, $\varphi_{\mu}$, $|A_t|$, $\varphi_{A_t}$, $|A_b|$, $\varphi_{A_b}$,
which determine the mass eigenvalues and mixing angles of the top and bottom 
squarks. In addition, in the chargino sector the $SU(2)$ gaugino mass
parameter
$M_2$ enters. The mass eigenvalues and mixing of the neutralino depend also on
the $U(1)$ gaugino mass parameter $M_1$ with phase 
$\varphi_\mathrm{U(1)}$ which, therefore, influences the partial 
decay widths of the
top and bottom squarks into neutralinos. 
The main parameter of the Higgs sector is the 
charged Higgs boson mass $m_{H^\pm}$, in addition to the parameters already 
introduced.

We calculate the partial decay widths in Born approximation. In some
cases the one-loop SUSY QCD corrections are important. The analyses of
\cite{stop1,eberlRealSf,mbrun2} suggest that a significant part of the
one-loop SUSY QCD corrections to the partial widths of 
$\ti t_i$ and $\ti b_i$ decays
(where the bottom Yukawa coupling is involved) can be
incorporated by using an appropriately corrected bottom quark mass. In
this spirit we calculate the tree-level widths of the $\st_i$ and
$\sb_i$ decays by using on-shell masses for the kinematic 
terms (such as phase space factors) and by taking running
$t$ and $b$ quark masses for the Yukawa couplings.
For definiteness we take
$m_t^\mathrm{run}(m_Z) = 150$~GeV,
$m_t^\textrm{\scriptsize on-shell} = 175$~GeV,
$m_b^\mathrm{run}(m_Z) = 3$~GeV and
$m_b^\textrm{\scriptsize on-shell} = 5$~GeV.
This approach leads to an ``improved'' Born approximation which takes
into account an essential part
of the one-loop SUSY QCD corrections to the $\st_i$ and $\sb_i$
partial decay widths and predicts their phase dependences more accurately
than the ``naive'' tree-level calculation.
In the calculation of the CP violating effects in the neutral Higgs
sector we take the program FeynHiggs-2.0.2 of \cite{feynhiggs}.

In the numerical analysis we impose as theoretical constraint
the approximate necessary condition for the tree-level 
vacuum stability \cite{Derendinger-Savoi}. Furthermore, as experimental 
constraints we take into account the mass bounds from LEP 
\cite{LEP} and $\Delta\rho(\ti t-\ti b) < 0.0012$ \cite{deltarho},
as well as 
 $2.0 \times 10^{-4} < B(b \to s \gamma) < 4.5 \times 10^{-4}$
 \cite{bsgamma} assuming the Kobayashi-Maskawa mixing also
for the squark sector.
For the calculation of the $b\to s\gamma$ width 
we use the formula
of \cite{Bertolini:1990if} including the O($\alpha_s$) corrections as
given in \cite{Kagan:1998ym}. We also take
$|M_1|=5/3\tan^2\theta_W M_2$.

\begin{figure}[t]
\setlength{\unitlength}{1mm}
\begin{center}
\begin{picture}(150,100)
\put(-50,-120){\mbox{\epsfig{file=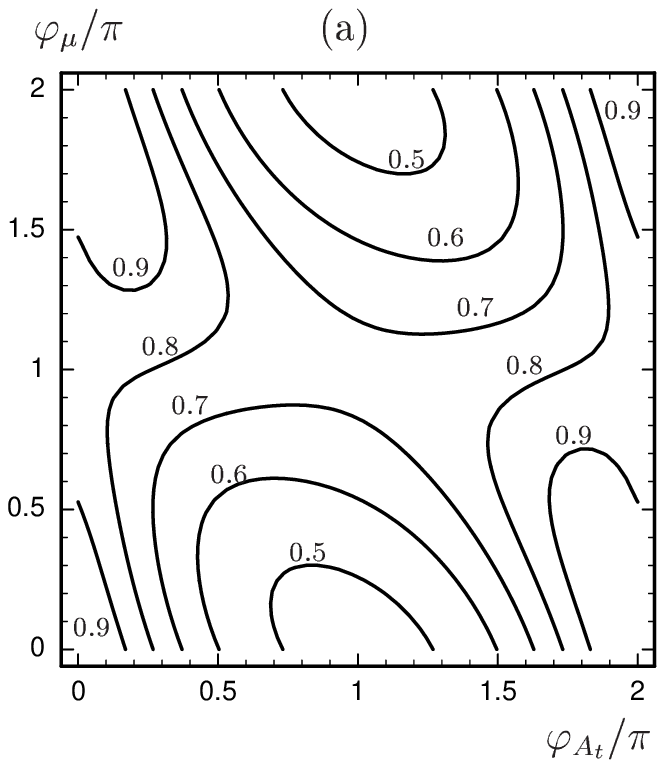,height=22.cm,width=20.cm}}}
\put(25,-124){\mbox{\epsfig{file=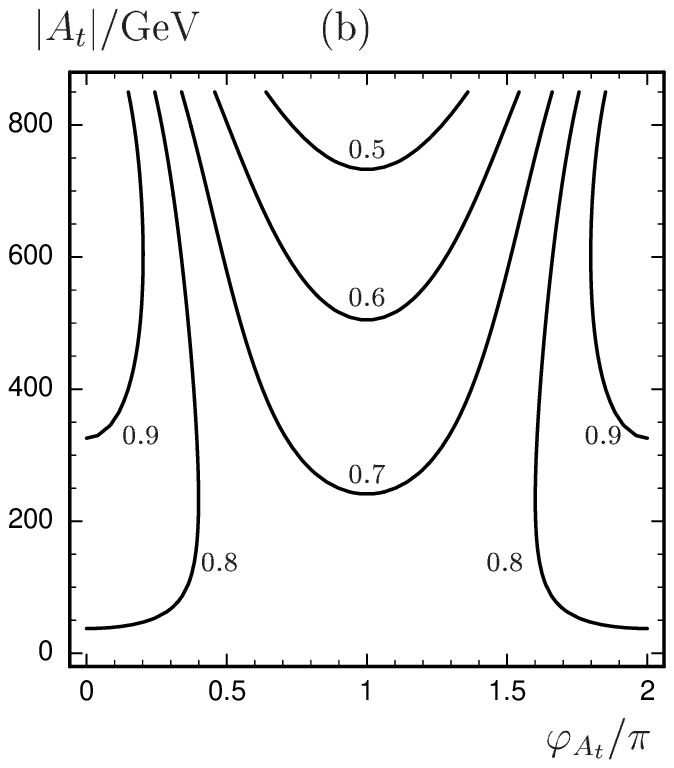,height=22.5cm,width=19.5cm}}}
\end{picture}
\end{center}
\vskip -2.5cm
\caption{Contours of $B(\tilde{t}_1 \to \tilde{\chi}^0_1 t)$ 
for $\tan\beta = 6$, $M_2=300$~GeV, $|\mu|=500$~GeV,
$\varphi_\mathrm{U(1)}=\varphi_{A_b}=0$,
$m_{\tilde{t}_1}=350$~GeV, $m_{\tilde{t}_2}=700$~GeV,
$m_{\tilde{b}_1}=170$~GeV, $m_{H^\pm}=600$~GeV,
with (a) $|A_t|=|A_b|=800$~GeV
and (b) $\varphi_\mu=0$, $|A_b|=|A_t|$, 
assuming $M_{\tilde{Q}}>M_{\tilde{U}}$.}\label{cpneut}
\end{figure}

As a first example we show 
in Fig.~\ref{cpneut} (a) the contour plot for
$B(\tilde{t}_1 \to \tilde{\chi}^0_1 t)$ as a function of $\varphi_{A_t}$
and $\varphi_\mu$ 
for $(m_{\tilde{t}_1}, m_{\tilde{t}_2}, m_{\tilde{b}_1})=
(350, 700, 170)$~GeV, 
$\tan\beta = 6$, $M_2=300$~GeV,
$|\mu|=500$~GeV, $|A_t|=|A_b|=800$~GeV,
$\varphi_\mathrm{U(1)}=\varphi_{A_b}=0$ and $m_{H^\pm}=600$~GeV,
assuming $M_{\tilde{Q}}>M_{\tilde{U}}$. 
For the parameters chosen the $\varphi_{A_t}$ dependence is
stronger than the $\varphi_\mu$ dependence.
The reason is that these phase dependences are caused mainly by the
$\tilde{t}_L$ -$\tilde{t}_R$ mixing term, where
the $\varphi_\mu$ dependence is suppressed by $\cot\beta$. The $\varphi_\mu$
dependence is somewhat more pronounced for $\varphi_{A_t} \approx \pi$
than for $\varphi_{A_t} \approx 0, 2\pi$.
In Fig.~\ref{cpneut} (b) we show the contour plot of 
$B(\tilde{t}_1 \to \tilde{\chi}^0_1 t)$ as a function of
$\varphi_{A_t}$ and $|A_t|$ for $\varphi_\mu = 0$ and
$|A_t|=|A_b|$. Clearly, the $\varphi_{A_t}$ dependence is stronger
for larger values of $|A_t|$.

\begin{figure}[t]
\setlength{\unitlength}{1mm}
\begin{center}
\begin{picture}(150,100)
\put(-40,-122){\mbox{\epsfig{file=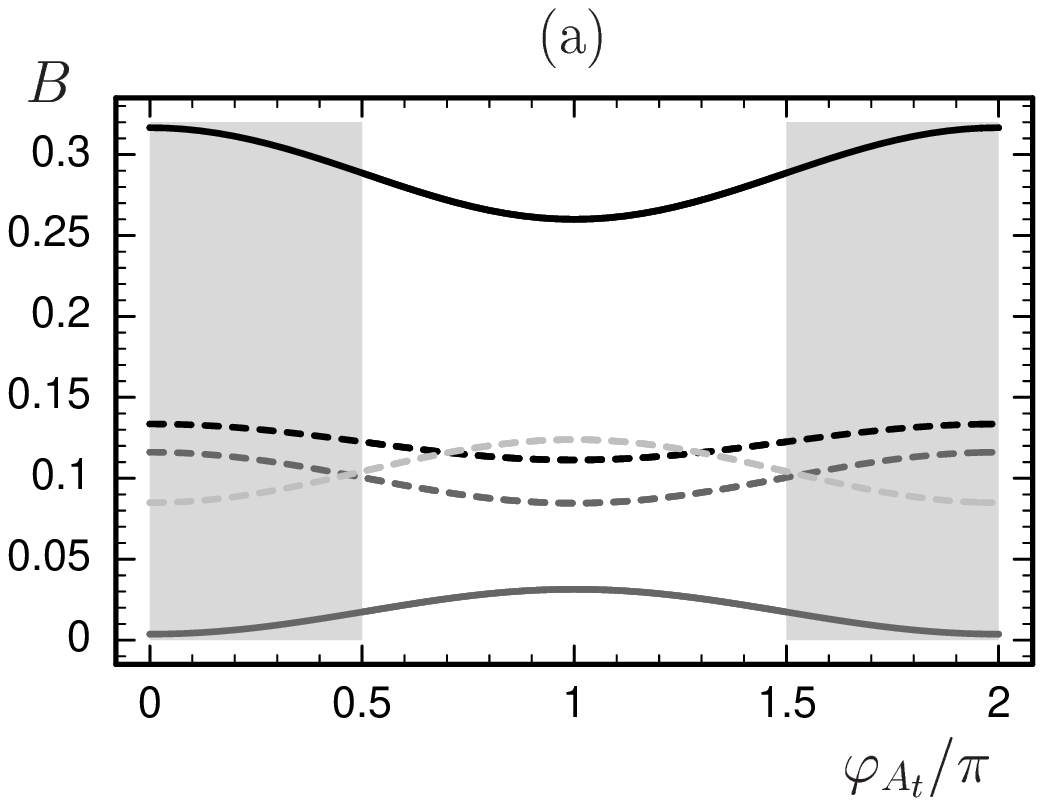,height=22.3cm,width=13.9cm}}}
\put(35,-124){\mbox{\epsfig{file=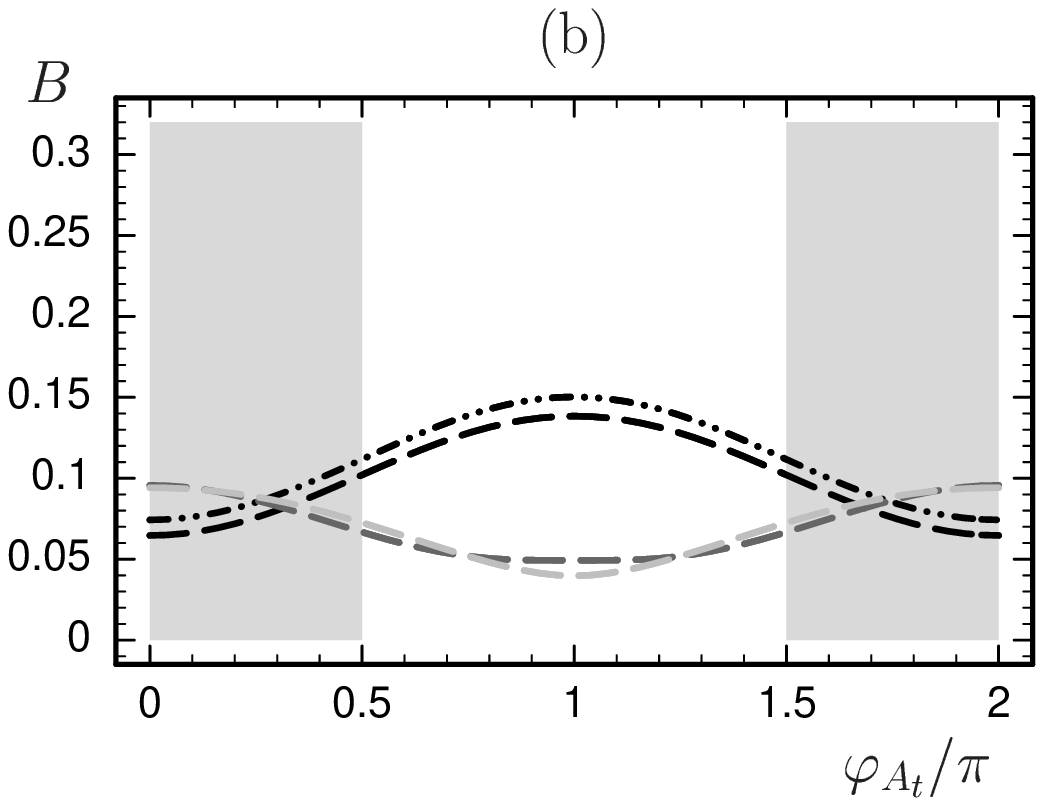,height=22.5cm,width=14cm}}}
\end{picture}
\end{center}
\vskip -2.5cm
\caption{\label{brphiAtSt2} $\varphi_{A_t}$ dependence of
branching ratios of the decays (a)
$\tilde{t}_2 \to \tilde{\chi}^+_{1/2} b$ (solid, black/gray),
$\tilde{t}_2 \to \tilde{\chi}^0_{2/3/4} t$ (dashed, black/gray/light gray)
and (b)
$\tilde{t}_2 \to Z \tilde{t}_1$ (dashdotdotted),
$\tilde{t}_2 \to H_{1/2/3} \tilde{t}_1$ (long dashed, black/gray/light
gray)
for $\tan\beta = 6$,
$M_2=300$~GeV, $|\mu| = 500$~GeV, $|A_b|=|A_t|=500$~GeV,
$\varphi_\mu=\varphi_\mathrm{U(1)}=\varphi_{A_b}=0$,
$m_{\tilde{t}_1}=350$~GeV, $m_{\tilde{t}_2}=800$~GeV,
$m_{\tilde{b}_1}=170$~GeV and $m_{H^\pm}=350$~GeV, assuming
$M_{\tilde{Q}}>M_{\tilde{U}}$. 
Only the decay modes with $B \gtrsim
1\,\%$ are shown.
The shaded areas mark the region excluded by the
experimental limit $B(b\to s \gamma) < 4.5 \times 10^{-4}$.}
\end{figure}

For the heavier top squark $\tilde{t}_2$ more decay channels are open. 
In Fig.~\ref{brphiAtSt2} (a) we show the branching ratios for 
$\tilde{t}_2 \to \tilde{\chi}^+_{1,2} b$ and
$\tilde{t}_2 \to \tilde{\chi}^0_{2,3,4} t$
as a function of $\varphi_{A_t}$ for 
$(m_{\tilde{t}_1}, m_{\tilde{t}_2}, m_{\tilde{b}_1})=
(350, 800, 170)$~GeV,
 $\tan\beta = 6$,
$M_2=300$~GeV, $|\mu| = 500$~GeV,
$|A_b|=|A_t|=500$~GeV,$\varphi_\mu=\varphi_\mathrm{U(1)}=\varphi_{A_b}=0$,
$m_{\tilde{t}_1}=350$~GeV, $m_{\tilde{t}_2}=800$~GeV,
$m_{\tilde{b}_1}=170$~GeV and $m_{H^\pm}=350$~GeV,
assuming $M_{\tilde{Q}}>M_{\tilde{U}}$.
The $\varphi_{A_t}$ dependence of 
$B(\tilde{t}_2 \to \tilde{\chi}^+_{1,2} b)$ is due to a direct phase
effect, which explains that the shape of 
$B(\tilde{t}_2 \to \tilde{\chi}^+_{1,2} b)$ is like 
($1\pm\cos\varphi_{A_t}$).
Also the phase dependence of the branching ratios into neutralinos
is mainly due to a direct phase effect.
$B(\tilde{t}_2 \to \tilde{\chi}^0_2 t)$ has a very weak phase dependence like
($10+\cos\varphi_{A_t}$).
In $\Gamma(\tilde{t}_2 \to \tilde{\chi}^0_3 t)$ the 
mixing phase enters, resulting in a shape like ($1+\cos\varphi_{A_t}$)
for the branching ratio. Similarly,
$B(\tilde{t}_2 \to \tilde{\chi}^0_4 t)$ behaves like
($1-\cos\varphi_{A_t}$).

In Fig.~\ref{brphiAtSt2} (b) we show the branching ratios for the
bosonic decays $\tilde{t}_2 \to Z \tilde{t}_1$ and 
$\tilde{t}_2 \to H_i \tilde{t}_1$ $(i=1,2,3)$ for the same parameter values
as above.
The shape of $B(\tilde{t}_2 \to Z \tilde{t}_1)$ is
like $(1-\cos\varphi_{A_t})$, which is caused by the
$\theta_{\tilde{t}}$ dependence of the corresponding coupling.
Quite generally, the
phase dependence of 
$\Gamma(\ti t_2\to H_k\ti t_1)$ 
is the result of a complicated
interplay among the phase dependences of the neutral Higgs boson 
masses, the top squark mixing matrix
elements,
the neutral Higgs mixing matrix elements 
and the direct top squark-Higgs couplings.
In the present example the $\varphi_{A_t}$ dependence of the partial
widths $\Gamma(\tilde{t}_2 \to H_{1,2,3} \tilde{t}_1)$ is mainly due
to the $\varphi_{A_t}$ dependence of the top squark mixing matrix 
and the squark-Higgs couplings, whereas the
$\varphi_{A_t}$ dependence of the neutral Higgs mixing matrix 
is less pronounced in this case. 

\begin{figure}[t]
\setlength{\unitlength}{1mm}
\begin{center}
\begin{picture}(150,100)
\put(-40,-122){\mbox{\epsfig{file=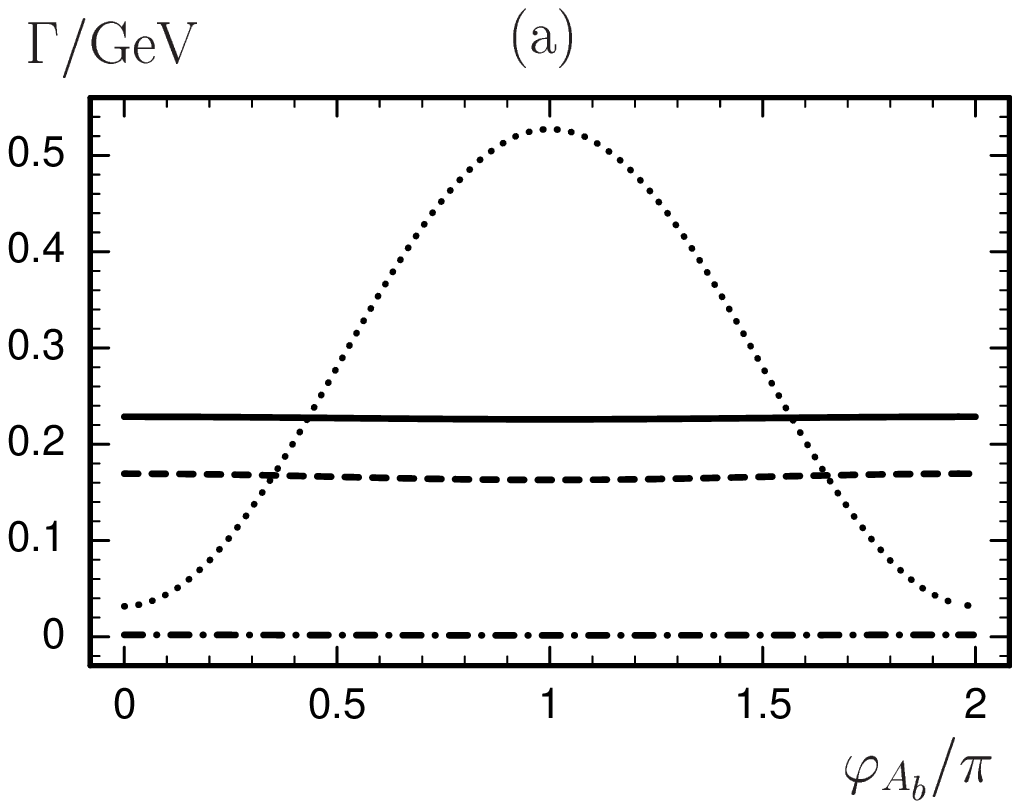,height=22.3cm,width=13.9cm}}}
\put(35,-124){\mbox{\epsfig{file=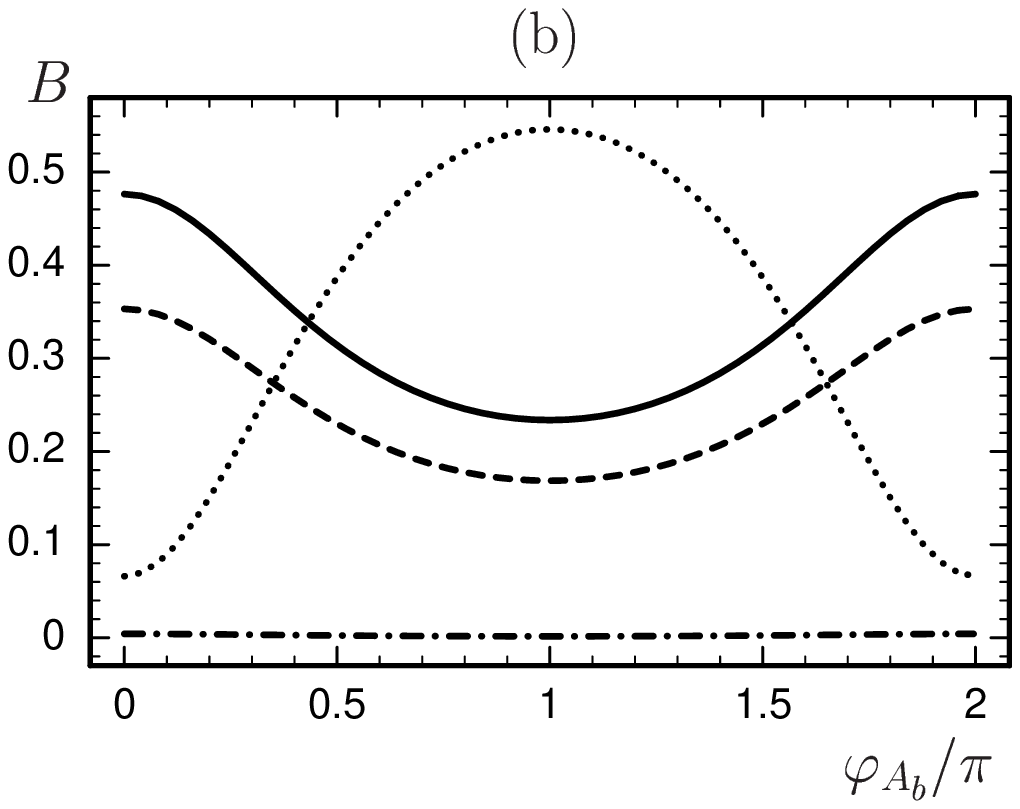,height=22.5cm,width=14cm}}}
\end{picture}
\end{center}
\vskip -2.5cm
\caption{\label{sbottomphiAb} $\varphi_{A_b}$ dependences of
(a) partial widths and (b) branching
ratios of the decays
$\tilde{b}_1 \to \tilde{\chi}^0_1 b$ (solid),
$\tilde{b}_1 \to \tilde{\chi}^0_2 b$ (dashed),
$\tilde{b}_1 \to H^- \tilde{t}_1$ (dotted) and
$\tilde{b}_1 \to W^- \tilde{t}_1$ (dashdotted)
for $\tan\beta = 30$, $M_2=200$~GeV, $|\mu| = 300$~GeV, $|A_b|=|A_t|=600$~GeV,
$\varphi_\mu=\pi$, $\varphi_{A_t}=\varphi_\mathrm{U(1)}=0$,
$m_{\tilde{b}_1}=350$~GeV, $m_{\tilde{b}_2}=700$~GeV,
$m_{\tilde{t}_1}=170$~GeV and $m_{H^\pm}=150$~GeV, 
assuming $M_{\tilde{Q}}>M_{\tilde{D}}$.}
\end{figure}

Coming now to the discussion of the decays of the bottom squarks
$\tilde{b}_{1,2}$, we show in 
Fig.~\ref{sbottomphiAb} the partial decay widths and
the branching ratios of $\tilde{b}_1 \to \tilde{\chi}^0_{1,2} b$,
$H^- \tilde{t}_1$, $W^- \tilde{t}_1$
as a function of $\varphi_{A_b}$ for 
$m_{\tilde{b}_1}=350$~GeV, $m_{\tilde{b}_2}=700$~GeV,
$m_{\tilde{t}_1}=170$~GeV, $\tan\beta = 30$, $m_{H^\pm}=150$~GeV,
$M_2=200$~GeV, $|\mu| = 300$~GeV,
$|A_b|=|A_t|=600$~GeV,
$\varphi_\mu=\pi$ and $\varphi_{A_t}=\varphi_\mathrm{U(1)}=0$, 
assuming $M_{\tilde{Q}}>M_{\tilde{D}}$.
In the region $0.5 \pi < \varphi_{A_b} < 1.5 \pi$ the decay
$\tilde{b}_1 \to H^- \tilde{t}_1$ dominates.
The $\varphi_{A_b}$ dependence of 
$\Gamma(\tilde{b}_1 \to H^- \tilde{t}_1)$ is due to the 
behaviour of the squark-Higgs coupling.
The partial decay widths $\Gamma(\tilde{b}_1 \to \tilde{\chi}^0_{1,2} b)$ are
almost $\varphi_{A_b}$ independent because the $\varphi_{A_b}$
dependence of the bottom squark mixing matrix 
nearly vanishes for $\tan\beta=30$.
Hence the $\varphi_{A_b}$ dependence of the branching ratios 
$B(\tilde{b}_1 \to \tilde{\chi}^0_{1,2} b)$ is caused by that
of the total decay width.
$\Gamma(\tilde{b}_1 \to W^- \tilde{t}_1)$ is suppressed because 
$\tilde{b}_1\sim \tilde{b}_R$ and $\tilde{t}_1\sim \tilde{t}_R$ 
in this scenario.

We have also estimated what accuracy can be expected in the
determination of the underlying MSSM parameters by
a global fit of the observables (masses, branching ratios and production
cross sections) measured at typical linear collider experiments with 
polarized beams. 
We have found that 
under favourable conditions the fundamental MSSM parameters except $A_{t,b}$
can be determined with errors of 1\,\% to 2\,\%, assuming an
integrated luminosity of 1~ab$^{-1}$. The parameter $A_t$ can be
determined within an error of 2 -- 3\,\% 
whereas the error of $A_b$ is likely to be of the
order of 50\,\%. More details can be found in \cite{stosbo}.

\section{CP Asymmetries in Sfermion Decays}

As the sfermions are scalar particles and they usually have 
two--body decay modes, it is not straightforward to construct a
CP sensitive asymmetry involving their decays. In the case of top
squarks one can use a three--body decay \cite{Bartl:2002hi}, 
for the other sfermions one has to consider cascade decays.
In our paper \cite{Bartl:2003ck} we have considered the decay chain
\begin{equation}
\ti f\to f\ti\chi^0_j\to f \ti\chi^0_1~Z\to 
f\ti\chi^0_1~\ell~\bar \ell 
~~(f\ti\chi^0_1~q~\bar q),
\end{equation}
where $\ell=e,\mu,\tau$, and $q$ denotes a quark. We have defined
a T-odd correlation for the leptonic decay 
\begin{equation}
O^{\ell}_{\rm odd}=
{\bf p}_{f}\cdot({\bf p}_{\ell}\times{\bf p}_{\bar \ell}),
\end{equation}
and for the hadronic decays as
\begin{equation}
O^{q}_{\rm odd}=
{\bf p}_{f}\cdot({\bf p}_{q}\times{\bf p}_{\bar q}),
\end{equation}
where ${\bf p}$ denotes the three-momentum of the corresponding
fermion.
We define the corresponding T-odd asymmetries as
\begin{equation}
{\mathcal A}_{\rm T}^{\ell,q}=
\frac{\Gamma(O^{\ell,q}_{\rm odd}>0)-\Gamma(O^{\ell,q}_{\rm odd}<0)}
{\Gamma(O^{\ell,q}_{\rm odd}>0)+\Gamma(O^{\ell,q}_{\rm odd}<0)}
\end{equation}
which by $CPT$ are also CP asymmetries. This CP asymmetry 
is similar to that proposed in \cite{Oshimo}, however, we 
calculate the asymmetry in the full phase space of the decay chain.
For the measurement of ${\mathcal A}_{\rm T}^{\ell}$
or ${\mathcal A}_{\rm T}^q$ it is necessary to
be able to distinguish between the charges
of ${\ell}^+$ and ${\ell}^-$ or $q$ and $\bar q$.
In the case $\ell=e,\mu,\tau$ this should be
possible experimentally on an event by event basis
at an $e^+e^-$ linear collider \cite{tdr}.
${\mathcal A}_{\rm T}^q$ will be measureable 
in the case of $q=c,b$, where flavour reconstruction 
is possible \cite{Damerell:1996sv}. 

As an example we have calculated
${\mathcal A}^\ell_{\rm T}$ for $\ti \tau_1$ decay, 
considering the decay chain
$\ti \tau_1 \to \tau~\ti\chi^0_2$, $\ti\chi^0_2 \to Z~\ti\chi^0_1$, 
$Z \to \ell~\bar \ell$, 
for $\ell=e,\mu,\tau$. As input parameters we have chosen
$m_{\ti\tau_1}=300$~GeV, $ m_{\ti\tau_2}=800$~GeV,
$\tan\beta =10$,
$|A_{\tau}|=1000$~GeV, $\varphi_{A_{\tau}}=0$,
$m_{A}=800$~GeV, $M_2=280$~GeV, using the GUT relation 
$|M_1|=5/3 \tan^2\theta_W M_2$

\begin{figure}[t]
\setlength{\unitlength}{1mm}
\begin{center}
\begin{picture}(150,100)
\put(-50,-120){\mbox{\epsfig{file=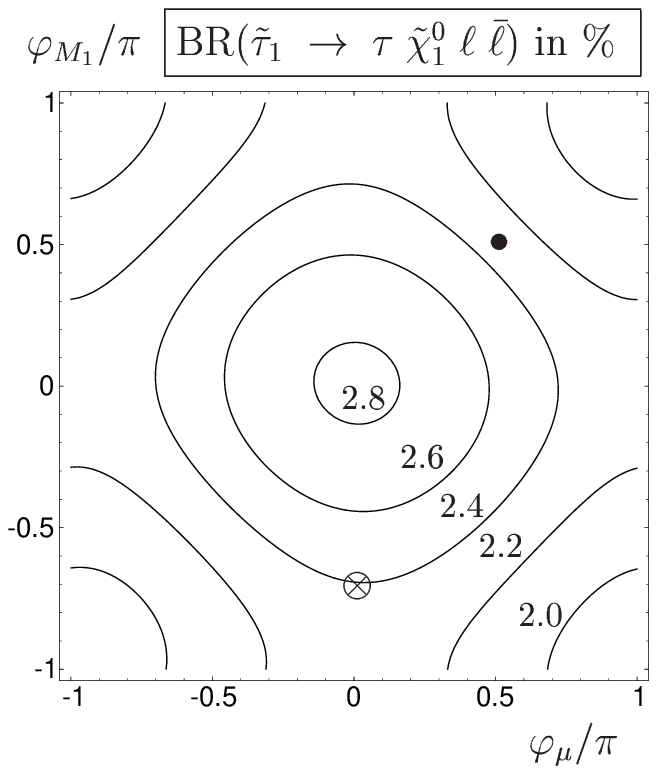,height=22.cm,width=20.cm}}}
\put(25,-124){\mbox{\epsfig{file=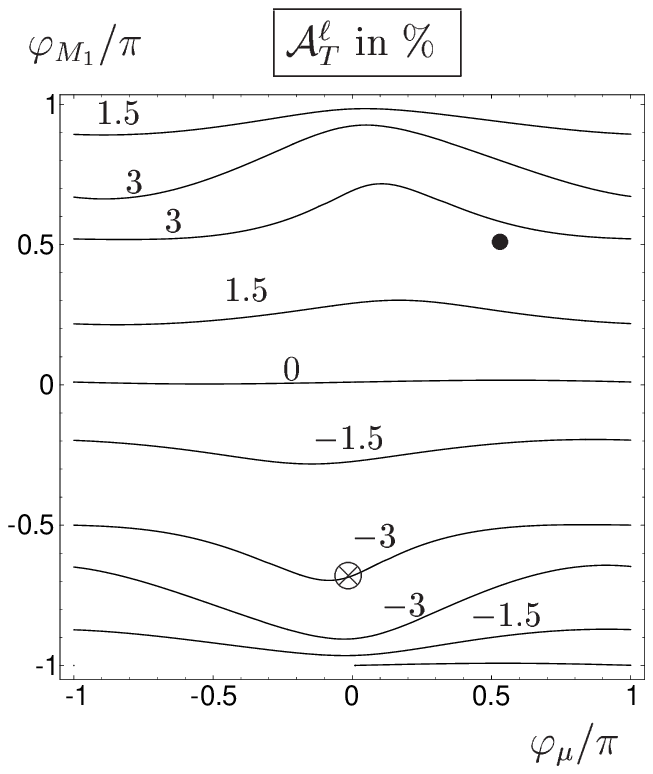,height=22.5cm,width=19.5cm}}}
\end{picture}
\end{center}
\vskip -2.5cm
\caption{
Contour lines of the branching ratio 
for $\ti\tau_1\to \ti\chi^0_1\tau \ell\bar\ell$ 
and asymmetry $A_T^\ell$ 
in the $\varphi_{ M_1}$-$\varphi_{\mu} $ plane 
for $|\mu|=300$~GeV and $M_2=280$~GeV, taking
$\tan\beta =10$, $A_{\tau}=1000$~GeV, 
$m_{\ti\tau_1}=300$~GeV, $ m_{\ti\tau_2}=800$~GeV
for $M_{\ti E} > M_{\ti L}$.
The point denoted by $\bullet$ is for the theoretical estimate 
of the necessary number of produced $\ti \tau_1$'s (see text).}
\label{stau2b}
\end{figure}

In Fig.~\ref{stau2b}a we show the contour lines for the branching ratio 
BR$(\ti \tau_1 \to \tau~\ti\chi^0_1~\ell~\bar \ell) =
{\rm BR}(\ti \tau_1 \to \tau~\ti\chi^0_2)
\times{\rm BR}(\ti\chi^0_2\to Z\ti\chi^0_1)\times
{\rm BR}(Z \to \ell~\bar \ell)$
in the $\varphi_{M_1}$-$\varphi_{\mu}$ plane
for $M_2=280$~GeV and $|\mu|=300$~GeV.
For BR$(\ti \tau_1 \to \tau~\ti\chi^0_1~\ell~\bar \ell)$ we always sum over
$\ell=e,\mu,\tau$.
We choose $M_{\ti E} > M_{\ti L}$ since in this case the 
$\ti \tau_1$-$\tau$-$\ti\chi^0_2$ coupling $|a^{\ti\tau}_{12}|$ is larger, 
which implies a larger branching ratio 
BR$(\ti \tau_1 \to \tau~\ti\chi^0_2)$
than for $M_{\ti E} < M_{\ti L}$. 
In a large region of the parameter
space we have BR$(\ti\chi^0_2\to Z\ti\chi^0_1) = 1$,
and we take BR$(Z \to \ell~\bar \ell)=0.1$.
In Fig.~\ref{stau2b}b we show 
the $\varphi_{M_1}$ and $\varphi_{\mu}$ dependence 
of ${\mathcal A}_{\rm T}^\ell$.
The value of ${\mathcal A}_{\rm T}^\ell$ depends stronger on $\varphi_{M_1}$,
than on $\varphi_{\mu}$. 
The sign of ${\mathcal A}_{\rm T}^\ell$ is essentially determined by the
sign of $\varphi_{M_1}$.

The asymmetries ${\mathcal A}_{\rm T}^{b(c)}$ can be calculated from
${\mathcal A}_{\rm T}^\ell$ by replacing the leptonic $Z$ coupling by the
$Z q \bar{q}$ coupling. This gives
\begin{equation}
{\mathcal A}_{\rm T}^{b(c)} \approx 6.3 (4.5) \times {\mathcal A}_{\rm T}^\ell.
\end{equation}

Based on our results for the asymmetry 
${\mathcal A}^{\ell }_{\rm T}$ in $\ti\tau_1\to\tau\ti\chi^0_2\to
\ti\chi^0_1\tau\ell^+ \ell^-$ and the branching ratio
we give a theoretical estimate of the number of produced
$\ti\tau_1$'s necessary to observe the T-odd asymmetry.
As an example we take the point denoted by $\bullet$ in Fig.~\ref{stau2b},
with $\varphi_{\mu}=\pi/2$ and $\varphi_{M_1} = \pi/2$.
For this point ${\rm BR} \approx 2.5\times 10^{-2}$
and $|{\mathcal A}_{\rm T}^\ell|\approx 3\times 10^{-2}$.
For the decay $\ti \tau_1\to b\bar b\ti\chi^0_1\tau$, on the other
hand, ${\rm BR} \approx 3.6\times 10^{-2}$ and 
$|{\mathcal A}_{\rm T}^b|\approx 1.9 \times 10^{-1}$.
In this example
the asymmeties ${\mathcal A}_{\rm T}^{\ell,q}$
should be measurable at an $e^+ e^-$ linear
collider  with $\sqrt{s}=800$~GeV and an integrated
luminosity of $500~fb^{-1}$
for $m_{\ti\tau_1}=300$~GeV.
It is clear that detailed Monte Carlo studies taking into
account background and detector simulations are necessary to get a more 
precise prediction of the expected accuracy.
For a Monte Carlo study on a T-odd observable in neutralino
production and decay see \cite{Choi:2003pq}.


\section*{Acknowledgements}
We thank J. W. F. Valle and the organizers of AHEP03 for creating 
an inspiring atmosphere at this conference.
This work is supported by the `Fonds zur F\"orderung der
wissenschaftlichen For\-schung' of Austria, FWF Projects No.~P13139-PHY
and No.~P16592-N02,
by the European Community's Human Potential Programme
under contract HPRN-CT-2000-00149
and by the 'Deutsche Forschungsgemeinschaft' (DFG) under 
contract Fr 1064/5-1.
O.K.\ is supported by
the EU Research Training Site contract HPMT-2000-00124,
by Spanish grants BFM2002-00345
and by the EU network Programme HPRN-CT-2000-00148.
W.P.\ has been supported by the Erwin Schr\"odinger fellowship 
No.~J2272 of the `Fonds zur F\"orderung der wissenschaftlichen 
Forschung' of Austria and partly by the Swiss 'Nationalfonds'.

\end{document}